# Optical LC-like resonances in high-index particles

Xiang-yang Liu[1,2] and Cheng-ping Huang[1,]*

[1]*Department of Physics, Nanjing Tech University, Nanjing 211816, China*

[2]*School of 2011, Nanjing Tech University, Nanjing 211816, China*

**Abstract**

Electric LC resonances, occurring in metallic circuits, govern the motion of free electrons or electric signals. In this paper, optical LC-like resonances in high-index particles (HIPs) or dielectric LC resonances have been studied, which involve bounded electrons and optical fields instead. The resonance effect is dominated by optical analogues of inductance and capacitance, which can be determined according to the electromagnetic energy bounded near the particle. The viewpoint of dielectric optical circuit with equivalent parameters facilitates the understanding of dielectric resonance effect. The result also provides a method for studying the optical properties of the HIPs.

*Email: cphuang@njtech.edu.cn

## 1. Introduction

In recent years, meta-materials and meta-surfaces, i.e., the artificially designed and constructed materials consisting of deep subwavelength building blocks, have attracted much interest [1-3]. These artificial materials may exhibit some exotic optical properties, such as optical magnetism, negative refraction, cloaking effect, etc. [4-8]. Usually, the unusual optical properties of meta-materials and meta-surfaces are associated with the resonance effects of the building blocks. For example, the metallic nanoparticles [e.g., the metallic nano-spheres, nano-antennas, split-ring resonators (SRR), etc.] own the localized plasmonic resonances [9], which may modify the dielectric response and light scattering significantly. The localized plasmonic resonance originates from the strong coupling between optical fields and free electrons of the metal particles. It was shown that, when the particle sizes are much smaller than the working wavelength, such plasmonic resonators can mimic the conventional LC circuits [10], thus establishing an LC resonance in the optical frequency range. The circuit viewpoint provides a simple and useful method for the study of metallic nanoresonators [11-15].

Unlike the metal particles, high-index particles (HIPs), e.g., the silicon spheres, cylinders, discs, etc., own the so-called Mie resonances [16-18]. By using the Mie resonances, interactions between light and matter can be manipulated as well. Compared with the metal particles, the HIPs contain only bounded electrons and there is no Ohmic loss. Consequently, the HIPs have been employed widely to construct the low-loss meta-materials and meta-surfaces [19-23]. Generally, the resonances of the HIPs can be revealed with the numerical simulations, which, however, are resource and time consuming. For regular dielectric particles such as the spheres, infinitely-long cylindrical and rectangular nanowires, the Mie theory and waveguide theory may be used [16, 19, 24, 25]. Moreover, the Mie theory has been generalized by F. Papoff et al. to the particles of arbitrary shape [26]. Nonetheless, these theories involve the complicated calculations (e.g., the special functions) and are also not easy to recover the underlying physics. Thus, an intuitive and efficient method for understanding and calculating the resonances of the HIPs is desirable.

In this paper, we suggest that the fundamental magnetic resonance of the HIPs can be investigated using the concept of optical LC-like resonance. Different from electric LC resonance, the optical LC-like resonance is closely correlated with the bounded electrons and optical or electromagnetic fields, without participation of metallic elements and conduction current. In theory, the HIP is characterized by two optical circuit parameters (called here the optical inductance and capacitance), which can be determined, respectively, with the magnetic flux (or magnetic energy of the displacement current) and electric displacement flux (or energy of the induced electric field). Accordingly, an LC-like resonance will be resulted for the electromagnetic fields localized near the dielectric particle. It should be emphasized that the optical LC-like resonance corresponds to the fundamental Mie resonance of the HIPs, just like the electric LC resonance describing the fundamental plasmonic resonance of the metal particles. The result may provide valuable insights for understanding the dielectric resonance of the HIPs.

The paper is organized as follows. In Sec. 2, the resonance effect is firstly deduced and verified from a simple ring-shaped HIP. The conclusion is then extended qualitatively in Sec. 3 to the HIP with arbitrary shape. Based on the conclusion of Sec. 3 and the iterative method, the LC-like resonance frequency of a cylindrical HIP with infinite and finite lengths was derived in Sec. 4. The obtained results have also been compared with the numerical simulations in both microwave and optical frequency range. And a short summary is presented in Sec. 5.

## 2. Dielectric rings: A simple example

To elaborate the above idea, we consider firstly a simple ring-shaped HIP with the subwavelength sizes [see Fig. 1(a)], which is usually difficult to deal with in the theory. The outer radius of dielectric ring is $R$, the radius of cross section is $r$ ($r<<R<<\lambda$), and the permittivity of the ring is $\varepsilon_d$ ($\varepsilon_d >> 1$; the dispersion of the HIP is neglected). The incident wave propagates with the magnetic field along the $z$ axis of the ring. Due to the electromagnetic induction effect, an induced electric field $E$ can be excited in the ring. The induced time-varying electric field leads to a displacement current [18], which flows along the ring and generates an oscillating magnetic field. The equation governing the induced electric field of the ring can be written as

$$U_d = -d\Phi_m/dt = E2\pi R, \tag{1}$$

where $U_d$ is the induced electromotive force and $\Phi_m$ is the magnetic flux generated by the displacement current. As the ring radius $r$ is much smaller than the outer radius $R$, here the induced electric field is treated as uniform in the ring.

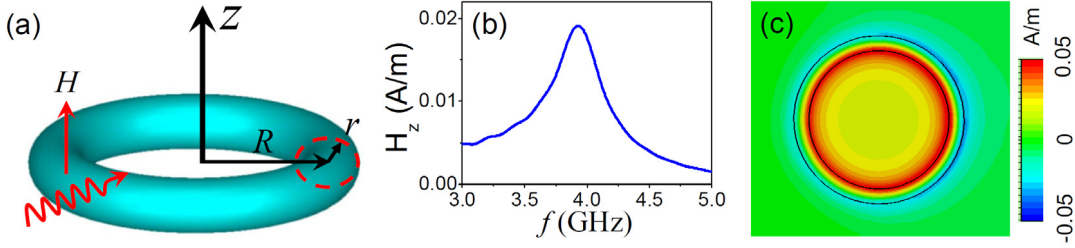

Fig. 1. (a) Schematic view of a ring-shaped HIP; (b) Magnetic field ($H_z$) at the ring center as a function of frequency; (c) $H_z$ distribution (xy plane) at the resonance frequency (3.93 GHz) of the ring. Here, R=10 mm, r=1 mm, and $\varepsilon_d = 100$.

For the circular dielectric ring studied here, two optical-circuit parameters can be defined and calculated as follows:

$$C_d = \frac{\Phi_d}{U_d} = \frac{\varepsilon_0 \varepsilon_d S}{2\pi R},$$
$$L_d = \frac{\Phi_m}{I_d} = \mu_0 R (\ln\frac{8R}{r} - \frac{7}{4}). \tag{2}$$

Here, $\Phi_d = DS$ is the electric displacement flux ($D = \varepsilon_0 \varepsilon_d E$ is the electric displacement in the ring, $S = \pi r^2$ is the ring cross-sectional area) and $I_d = d\Phi_d/dt$ is the displacement current of the ring. In our case, $L_d$ equals the inductance of a metal ring of the same shape and sizes [27].

With the use of Eq. (2), Eq. (1) can be rewritten as

$$\frac{d^2}{dt^2}\Phi_d + \frac{1}{L_d C_d}\Phi_d = 0. \tag{3}$$

Equation (3) is just similar to that of the electric LC circuit, with the electric displacement flux $\Phi_d$ corresponding to the quantity of free charges $Q$. The above result also indicates that there are optical analogues of the inductance and capacitance existing for the dielectric ring. To be distinct from the circuit parameters associated with the conduction current and free charges, the optical-circuit parameters introduced here may be termed as the optical inductance and capacitance, as they are linked to the optical fields bounded near the dielectric particle and irrelevant with the free charges. The optical inductance ($L_d$) correlates with the magnetic flux or magnetic-field energy ($W_m = L_d I_d^2 / 2$) of the displacement current and depends on the dielectric particle sizes. Instead, the optical capacitance ($C_d$) correlates with the electric displacement flux or energy of the induced electric field ($W_e = \Phi_d^2 / 2C_d$) and depends on the permittivity as well as the sizes of the particle.

Consequently, an LC-like resonance will be induced in the dielectric ring for the localized optical fields. Also, to be distinct from the electric LC resonance, such a resonance may be called the optical LC-like resonance or dielectric LC resonance. With Eqs. (3) and (2), the resonance frequency, $\omega_0 = 1/\sqrt{L_d C_d}$, can be deduced easily as

$$\omega_0 = \frac{c}{n_d r} \sqrt{\frac{2}{\ln(8R/r) - (7/4)}}, \qquad (4)$$

where $n_d = \sqrt{\varepsilon_d}$ is the refractive index of the dielectric ring. Note that, as $\Phi_d \propto D \approx P$ ($\varepsilon_d \gg 1$) and the polarization $P = -Ne\Delta r$ ($N$ and $\Delta r$ represents, respectively, the density and displacement of bounded electrons), the motion of bounded electrons in the dielectric ring will satisfy the similar equation as Eq. (3). Accordingly, the resonance frequency of the bounded electrons is also determined by Eq. (4) rather than by their inherent resonance frequency (this means we can modify the resonance frequency of the bounded electrons in atoms with dielectric structures). The LC-like resonance is typically accompanied by an enhanced magnetic near field, because of the resonant displacement current inside the particle.

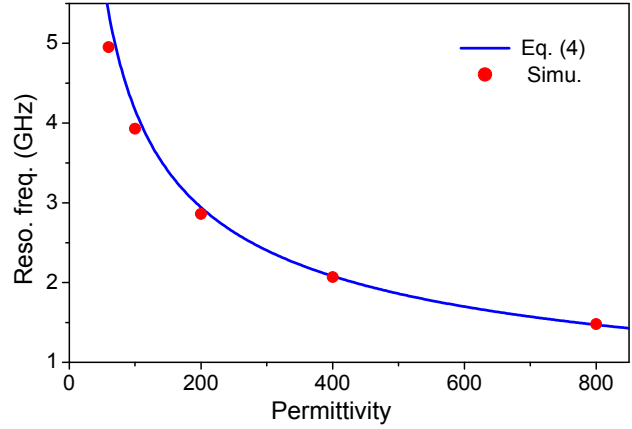

*Fig. 2. Relationship between the resonance frequency and permittivity of the ring-shaped HIP: the solid line represents the analytical results [Eq. (4)] and the circles the numerical simulations. Here, R=10 mm and r=1 mm.*

To confirm the above idea, numerical simulations based on the finite-difference time-domain method have been carried out [28]. In the simulation, open boundary conditions along the x-, y-, and z-axes were used, the incident field was set as $E_i = 1$ V/m (or $H_i = 2.65 \times 10^{-3}$ A/m), and a probe was placed at the ring center to record the magnetic field. Here, we take the microwave regime as an example (the conclusion can be extended to the optical range). Without loss of generality, the geometrical and physical parameters of the dielectric ring are set as *R*=10 mm, *r*=1 mm, and $\varepsilon_d = 100$ (such a permittivity is available for the high dielectric constant microwave ceramics [29]). Figure 1(b) presents the simulated magnetic field $H_z$ at the ring center as a function of frequency. One can see that, at the frequency 3.93 GHz, a peak of $H_z$ appears in the spectrum. The peak position is close to the analytical result (4.16 GHz) given by Eq. (4). The maximal $H_z$ at the ring center is 0.019 A/m, which is 7.2 times larger than the incident magnetic field. Figure 1(c) also plots the $H_z$ distribution on the xy plane at 3.93 GHz, with the result similar to that of a metallic SRR (the magnetic field near the ring surface is even stronger than that of the ring center). The enhanced magnetic field reveals the LC-like resonance of the dielectric ring. In addition, Fig. 2 shows the relationship between the resonance frequency and permittivity of the dielectric ring, where the ring sizes are fixed (*R*=10 mm and *r*=1 mm). When $\varepsilon_d$ varies from 60 to 800, the simulated resonance frequency decreases from 4.95 to 1.48 GHz (the circles), approaching the analytical results varying from 5.37 to 1.47 GHz (the solid line). The agreement between theory and simulation becomes better with the increase of $\varepsilon_d$, as the subwavelength condition ($R<<\lambda$) can be better satisfied for a higher refractive index of the ring.

### 3. Arbitrary particles: A general discussion

We can generalize the above discussions from the simple dielectric ring to an arbitrarily-shaped HIP. For the purpose, the fields are written in terms of the variable separation as $\vec{D}(r,t) = \vec{d}(r)D_0(t)$ and $\vec{H}(r,t) = \vec{h}(r)H_0(t)$, where $\vec{d}(r)$, $\vec{h}(r)$ are the spatial distribution functions of the fields and $D_0(t)$, $H_0(t)$ represent the time-varying terms. According to the Maxwell equation $\nabla \times \vec{H} = \partial \vec{D}/\partial t$, we set

$$\nabla \times \vec{h}(r) = \vec{d}(r),$$
$$H_0(t) = \frac{\partial D_0(t)}{\partial t}. \tag{5}$$

Then the electromagnetic energy bounded by the HIP can be expressed as

$$W = \int_{\Delta V} \frac{D^2}{2\varepsilon_0 \varepsilon_d} dV + \int_{NR} \frac{1}{2}\mu_0 H^2 dV$$
$$= \frac{D_0^2(t)}{2C_d} + \frac{1}{2}L_d H_0^2(t), \tag{6}$$

where the equivalent optical-circuit parameters are defined as

$$C_d = \frac{\varepsilon_0 \varepsilon_d}{\int_{\Delta V} |\vec{d}(r)|^2 dV},$$
$$L_d = \mu_0 \int_{NR} |\vec{h}(r)|^2 dV. \tag{7}$$

The integrals in Eqs. (6) and (7) run over the volume ($\Delta V$) or the near region (NR) of the HIP, where the electromagnetic fields are highly confined (The electric field energy, $\propto \varepsilon_d E^2$, is mainly localized in the high-index region, i.e., the inside of the particle. In contrast, as the permeability is uniform in the space, the localized magnetic-field energy can spill over the particle to the near region). Note that, because of the arbitrary shape of the HIP, here the expression of energy and the definition of optical capacitance and inductance are slightly different from that given previously, but this does not influence the conclusion.

Compared with the metallic particles, there is no Ohmic loss in the dielectric resonators. Moreover, under the LC-like resonance, the radiation loss of the HIP is very small. This means the resonance may have a high Q factor. Considering these facts, we can neglect the energy loss of the HIP. With the energy conservation ($\partial W / \partial t = 0$) and Eq. (5), we obtain

$$\frac{d^2}{dt^2}H_0(t) + \frac{1}{L_d C_d}H_0(t) = 0. \tag{8}$$

Equation (8) suggests that the arbitrarily-shaped HIP can present a harmonic oscillation, similar to the free electric LC circuit.

With Eqs. (7) and (5), the LC-like resonance frequency of an arbitrary HIP, $\omega_0 = 1/\sqrt{L_d C_d}$, is derived as

$$\omega_0 = \frac{c}{n_d}\sqrt{\frac{\int_{\Delta V}|\nabla \times \vec{h}(r)|^2 dV}{\int_{NR}|\vec{h}(r)|^2 dV}}. \tag{9}$$

Equation (9) indicates that the resonance frequency scales universally as the inverse of the particle refractive index $n_d$, regardless of the shape and sizes of the HIP.

In principle, once the spatial distribution function of magnetic near field $\vec{h}(r)$ is given or deduced for an arbitrary HIP, the equivalent optical-circuit parameters and the LC-like resonance frequency can then be determined [the amplitude of $\vec{h}(r)$ will not change the resonance frequency in Eq. (9)]. Actually, it is difficult to deduce analytically the detailed near-field distribution for an arbitrary particle (in this case, numerical simulation can present some useful information instead). Nonetheless, for a regular dielectric particle, Eq. (9) may provide a useful method for calculating the resonance frequency. To demonstrate this point, a cylindrical HIP with infinite and finite lengths will be taken as an example in the following.

## 4. The LC-like resonances of dielectric cylinders

We firstly consider an infinitely-long dielectric cylinder, as shown in Fig. 3(a). The cylinder has a radius $R$ ($R \ll \lambda$) and the incident wave propagates with the magnetic field along the $z$ axis. Such a cylindrical resonator can present a strong resonance, as demonstrated numerically in Figs. 3(b) and 3(c) (for the cylinder with $R=5$ mm and $\varepsilon_d = 60$, for example, a magnetic resonance appears at 2.90 GHz; the resonant magnetic field on the central axis attains 0.09 A/m, which is 34 times larger than the incident field). To solve the magnetic field distribution in the cylinder, an iterative method will be adopted here for simplicity. Still, the real electromagnetic fields are written as $\vec{E}(r,t) = \vec{e}(r)E_0(t)$ and $\vec{H}(r,t) = \vec{h}(r)H_0(t)$. With Maxwell equations $\nabla \times \vec{E} = -\mu_0 \partial \vec{H}/\partial t$, $\nabla \cdot \vec{E} = 0$ and boundary conditions, the electric field distribution can be expressed as a function of the magnetic one, i.e., $e(r) = F_1[h(r)]$. Similarly, with $\nabla \times \vec{H} = \varepsilon_0 \varepsilon_d \partial \vec{E}/\partial t$, $\nabla \cdot \vec{H} = 0$ and boundary conditions, we can also write $h(r) = F_2[e(r)]$. Combining the equations above gives

$$h(r) = F_2[F_1[h(r)]] \equiv F[h(r)]. \tag{10}$$

To solve Eq. (10), we consider a trial solution $h_0(r)$ with the same symmetry as that of $h(r)$. It can be proved mathematically that the function sequence

$$h_0(r),\ F[h_0(r)],\ F[F[h_0(r)]],\ \dots \tag{11}$$

will converge at $h(r)$. This method is applied below to determine the magnetic field distribution in the cylindrical resonators.

Assuming that a magnetic field along the z axis, $\vec{H}_i = \vec{e}_z h_i(r) H_0(t)$, is present in the dielectric cylinder. According to the Faraday's law, an induced electric field $\vec{E}_{i+1} = \vec{e}_\theta E_{i+1}(r,t)$ can be generated ($r \leq R$):

$$E_{i+1}(r,t) = -\frac{\mu_0}{r}\frac{dH_0(t)}{dt}\int_0^r h_i(x)x\,dx. \tag{12}$$

The induced oscillating electric field will produce a displacement current, which circulates along the azimuthal direction of the cylinder. The displacement current, in turn, causes the magnetic field pointing at the *z* axis. With the magnetic loop theorem, the resulting magnetic field is written as

$$H_{i+1}(r,t) = H_{i+1}(0,t) - \int_0^r \frac{\partial}{\partial t} D_{i+1}(y,t) dy. \tag{13}$$

By using Eq. (12), we have

$$H_{i+1}(r,t) = H_{i+1}(0,t) + \frac{\varepsilon_d}{c^2} \frac{d^2 H_0(t)}{dt^2} \int_0^r \frac{1}{y} \left( \int_0^y h_i(x) x dx \right) dy. \tag{14}$$

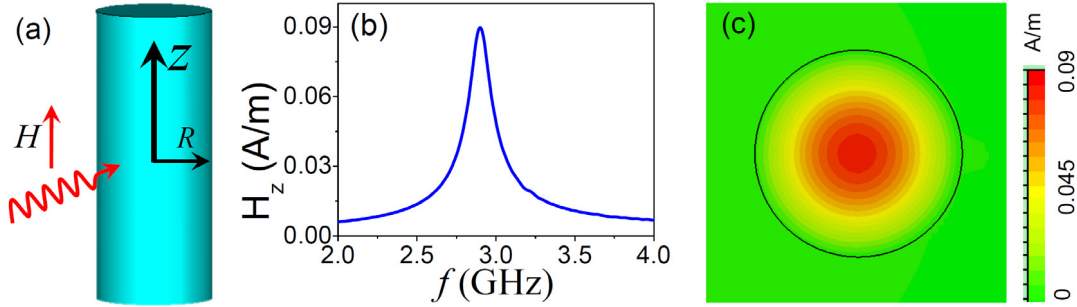

*Fig. 3. (a) Schematic diagram of an infinitely-long cylindrical HIP. The incident wave propagates with magnetic field along the z axis; (b) Simulated dependence of magnetic field $H_z$ (at the center of cylinder) on the frequency, and (c) $H_z$ distribution (xy plane) at the resonance frequency (2.90 GHz). Here, R=5 mm and $\varepsilon_d = 60$.*

Since we are treating the localized resonance, a boundary condition $H_{i+1}(R,t) = 0$ can be used [The infinitely-long cylinder carrying displacement current is similar to the infinitely-long solenoid carrying conduction current, with the magnetic field being zero outside the cylinder or solenoid. Then $h(r) = 0$ when $r \geq R$]. Therefore,

$$H_{i+1}(r,t) = \frac{\varepsilon_d}{c^2} \frac{d^2 H_0(t)}{dt^2} [P_i(r) - P_i(R)], \tag{15}$$

where

$$P_i(r) = \int_0^r Q_i(y) dy,$$
$$Q_i(y) = \frac{1}{y} \int_0^y h_i(x) x dx. \tag{16}$$

For time-harmonic fields, the magnetic-field distribution function in the cylinder can thus be obtained with Eq. (15) as

$$h_{i+1}(r) = k^2[P_i(R) - P_i(r)]. \tag{17}$$

Here $k = n_d \omega / c$ is the wavevector in the cylindrical resonator.

We set the initial distribution function of the exciting magnetic field as $h_0(r) = 1$ (the incident field is taken to be uniform in the cross section of the cylinder, as the wavelength is much larger than the radius). With Eqs. (16) and (17), we deduce that

$$\begin{aligned} h_1(r) &\propto 1 - \frac{r^2}{R^2}, \\ h_2(r) &\propto 1 - \frac{4r^2}{3R^2} + \frac{r^4}{3R^4}, \\ h_3(r) &\propto 1 - \frac{27r^2}{19R^2} + \frac{9r^4}{19R^4} - \frac{r^6}{19R^6}, \\ &\ldots. \end{aligned} \tag{18}$$

Thus, the magnetic-field distribution function in the cylinder appears as a polynomial expansion (the more terms in the expansion, the higher accuracy the distribution function).

By substituting Eq. (18) into Eq. (9) [$\vec{h}(r) = h(r)\vec{e}_z$], the resonance frequency of the dielectric cylinder is derived as

$$\omega_0 = \alpha \frac{c}{n_d R}. \tag{19}$$

Here $\alpha$ is a constant. For $h_1(r)$, $h_2(r)$, and $h_3(r)$, $\alpha$ can be determined as 2.450, 2.406, and 2.405, respectively. Thus, with only two round of iteration, a convergence of the resonance frequency can be approached. Hereafter, the constant $\alpha$ will be chosen as $\alpha = 2.40$ for the calculation. The result is in accordance with that obtained with the Mie theory [19]. Equation (19) suggests that the resonance wavelength of an infinitely long cylinder scales with the refractive index and radius of the cylinder.

When the dielectric cylinder is of the finite length $l$, there is also a standing-wave pattern along the $z$ direction. In this case, the wavevector in the cylinder satisfies $\varepsilon_d \omega^2 / c^2 = k_r^2 + k_z^2$. The standing-wave condition requires that $2k_z l = 2\pi m$ (where $m$ is an integer; as an approximation, the phase shift at the ends of the cylinder is neglected), with the $z$ component of the wavevector being $k_z = \pi m / l$. If the cylinder is infinitely long, then $k_z$ approaches zero and the cylinder resonates at $\omega_0$ [Eq. (19)]. Correspondingly, the radial wavevector is $k_r = n_d \omega_0 / c$. With the use of equation $\varepsilon_d \omega^2 / c^2 = k_r^2 + k_z^2$ and the standing-wave condition, the resonance frequency of the finite cylinder can be expressed approximately as

$$\omega_l = \sqrt{\omega_0^2 + (\frac{\pi c}{n_d l})^2}. \tag{20}$$

Here, only the lowest-order mode ($m$=1) has been considered. By substituting Eq. (19) into Eq. (20), we get

$$\omega_l = \frac{c}{n_d}\sqrt{\frac{\alpha^2}{R^2} + \frac{\pi^2}{l^2}}. \tag{21}$$

Hence, the smaller the cylinder length is, the higher the resonance frequency.

The theoretical results have been compared with the numerical simulations in the microwave band, as shown in Fig. 4. Figure 4(a) presents the resonance frequency of an infinitely long cylinder as a function of the radius $R$, where the permittivity is set as $\varepsilon_d = 60$. When $R$ increases from 2 to 10 mm, the resonance frequency predicted by Eq. (19) decreases from 7.34 to 1.48 GHz, which agrees with the simulation varying from 7.27 to 1.45 GHz. The deviation between theory and simulation is about 1.0~2.1%. Figure 4(b) plots the dependence of resonance frequency on $\varepsilon_d$ of an infinitely long cylinder, where the radius is fixed as $R = 10$ mm. The larger is the permittivity, the smaller the resonance frequency and the better the agreement between theory and simulation. In addition, Fig. 4(c) maps the resonance frequency of a cylinder with the finite length, where $R$=10 mm, $\varepsilon_d = 60$, and the cylinder length $l$ varies from 20 to 100 mm. In this case, the analytical results obtained from Eq. (21) are also close to the numerical simulations. A slightly larger deviation, 2.4~5.8%, can be attributed to the minor phase shift of the fields at the ends of the cylinder, which has been neglected in the theoretical treatment [noticing that we use Eq. (21) rather than Eq. (9) to calculate the resonance frequency, with the benefit of avoiding calculation of the detailed field distributions of the finite-length cylinder; We expect that, if the near field distributions of the finite structure is addressed, Eq. (9) can provide better agreement with the simulation].

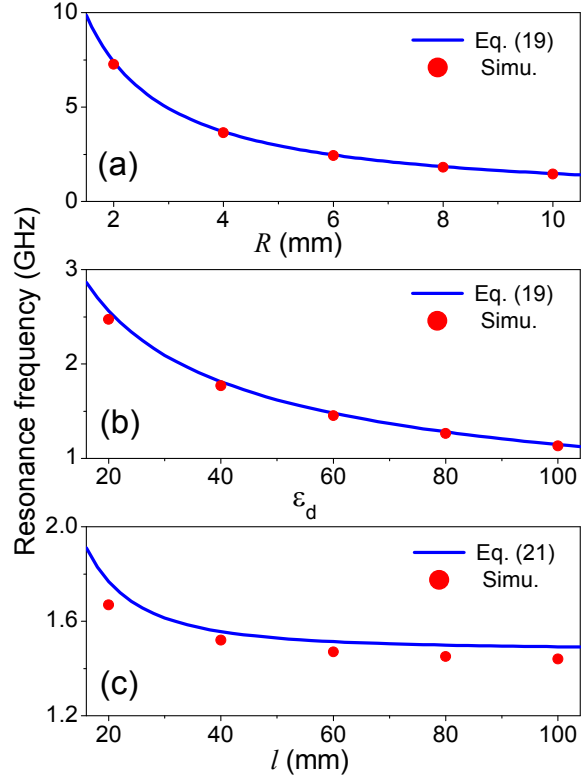

*Fig. 4. Resonance frequency $f_0$ of the cylindrical HIP in the microwave band: (a) $f_0$ versus R for an infinitely-long cylinder ($\varepsilon_d = 60$); (b) $f_0$ versus $\varepsilon_d$ for an infinitely-long cylinder (R=10 mm); (c) $f_0$ versus l for a cylinder of finite length ($\varepsilon_d = 60$, R=10 mm).*

The result can be extended to the optical frequency range as well. In the optical range, the high-index material silicon, with a permittivity of $\varepsilon_d = 12$, has been widely employed [20]. Here, by using the analytical formulas [Eqs. (19) and (21)] and numerical simulations, the resonance wavelength of a silicon cylinder with infinite and finite length has been calculated. Figure 5(a) plots the relationship between the resonance wavelength and radius R of an infinitely-long silicon cylinder. When R varies from 40 to 100 nm, the resonance wavelength obtained with Eq. (19) (the solid line) grows from 363 to 907 nm and the numerical value (the circles) grows from 380 to 950 nm. Moreover, for the silicon cylinder of finite length, the variation of resonance wavelength as a function of length *l* is shown in Fig. 5(b), where the radius is fixed as *R*=100 nm. The analytical results [Eq. (21)] still approach the simulations (the deviation is about 1.8~3.6%). These results demonstrate the applicability of theory in the optical frequency band.

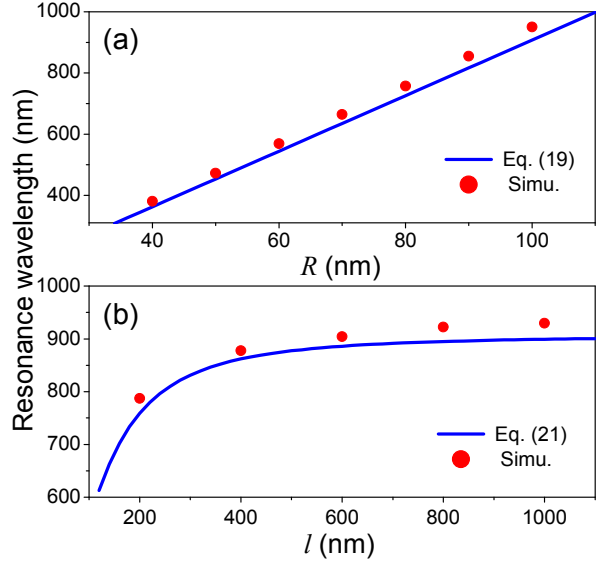

*Fig. 5. Resonance wavelength $\lambda_0$ of the cylindrical resonator (silicon with $\varepsilon_d = 12$) in the optical range: (a) $\lambda_0$ versus R for an infinitely-long cylinder; (b) $\lambda_0$ versus l for a cylinder of finite length (R=100 nm).*

### 5. Conclusions

In summary, an optical LC-like resonance in the HIPs has been suggested and investigated. In such effect, the optical or electromagnetic fields oscillate in a manner similar to the electric LC resonance, but the metallic elements and conduction current are excluded. The resonance frequency is governed by the optical inductance and capacitance, which can be derived from the electromagnetic energy bounded near the HIP. The obtained LC-like resonance frequency scales universally as the inverse of particle refractive index and depends strongly on the particle sizes. The theory has been employed successfully to calculate the magnetic resonance frequency of the ring-shaped and cylindrical resonators. The result provides an intuitive and simple way for understanding and studying the fundamental resonance of the HIPs, especially the dielectric particles with the rotational symmetry (including the dielectric rings with circular or rectangular cross section, hollow or solid dielectric cylinders, and multilayer cylinders, etc.).

The potential usefulness of the dielectric LC model lies in several aspects. Firstly, the magnetic resonance of the metallic SRRs and sandwiches plays a crucial role in realizing optical magnetism and negative refraction [1, 4, 5]. The high-index dielectric rings and cylinders, owning strong magnetic response and negligible dielectric loss, are ideal candidates for constructing three-dimensional metamaterials [19]. Our model will be helpful for the design of dielectric microstructure and derivation of the effective permeability. Secondly, in quasi-two-dimensional metasurfaces, the coupling between bright and dark modes, which are of larger contrast of dissipation loss, can mimic electromagnetically-induced transparency (EIT) effect [30]. The magnetic resonance of metallic rod pairs or SRR has been widely used as the subradient mode, because of its weak far-field radiation [30-32]. The dielectric LC resonance may be valuable for designing the high-Q EIT effect, replacing the lossy metallic SRRs with high-index dielectric rings [33, 34]. Thirdly, the model can be employed easily to calculate the magnetic dipole moment, scattering, absorption, and extinction spectra (considering the dielectric loss) of the single HIPs as well as the magnetic-dipole coupling between two or more HIPs (e.g., the dielectric rings). This is useful for tailoring the dielectric particle radiation and studying the optically-induced force of the dielectric particles (working under the magnetic resonance). And lastly, the toroidal dipole (TD) modes in metallic or dielectric particles have received much attention recently [35-37]. Although the proposed theory is suitable for the fundamental magnetic mode of the HIPs, some extensions are still possible. For example, by considering the conservation of electromagnetic energy bounded in the particle, the TD resonance frequency for some specific geometry (the dielectric ring or toroid) could be revealed.

**Acknowledgement**

This work was supported by the National Natural Science Foundation of China (Grant No. 12174193).